\documentstyle{elsart}
\begin{document}
\begin{frontmatter}
\title{Symmetry Tests of the Electroweak Interaction\\
from Muon Capture on \nuc{3}{He}}

\author{Jan Govaerts}
\address{Institut de Physique Nucl\'eaire\\
Universit\'e catholique de Louvain,\\
2, Chemin du Cyclotron\\
B-1348 Louvain-la-Neuve, Belgium\\
{\tt govaerts@fynu.ucl.ac.be}}

\begin{abstract}

Precision measurements in muon capture on \nuc{3}{He} to the
triton channel provide for interesting tests of the charged electroweak 
interaction, whether within the Standard Model or beyond it.
Based on the statistical capture rate and the triton asymmetry, both
the object of recent and on-going experiments using polarised
\nuc{3}{He} targets, examples of such
tests are presented, and shown to lead to stringent complementary constraints
for which further dedicated experimental as well as theoretical
efforts are required.\\


\vspace{40pt}

\begin{center}
Invited contribution to the 7$^{\mathrm th}$ RCNP International Workshop\\
HELION97: from Quark to Life\\
Polarized \nuc{3}{He} Beams and Gas Targets and Their Applications\\
Kobe, Japan, January 20--24, 1997
\end{center}

\vspace{165pt}

\noindent UCL-IPN-97-P01\\
January 1997

\end{abstract}

\end{frontmatter}

\clearpage

\section{Introduction}
\label{Sect1}

Ever since the discovery of parity violation in the weak interactions,
muon capture on nuclei has remained a topic of wide ranging 
interest\cite{Muk77}. Given the presently available high intensity muon beams
as well as new experimental techniques and particle detectors, 
enabling new measurements of much improved precision,
this interest has been revived in recent years\cite{LEMS93}.

Essentially two types of observables are presently available
in the case of muon capture on \nuc{3}{He} to the triton channel, 
$\mu^-\,+\,^3{\mathrm{He}}\,\rightarrow\,\nu_\mu\,+\,^3{\mathrm{H}}$,
namely the statistical muon capture rate $\lambda_{\mathrm stat}$
and the triton asymmetry $A_{\mathrm v}$ for a polarised
initial $\mu^--$\nuc{3}{He} state. The latter measurement is in progress,
and only a preliminary result is available so far\cite{Cum95}.
In the former case however, a recent
high precision measurement has reached the following value\cite{Acker97}, 
precise to 0.3\%,
\begin{equation}
\lambda^{\mathrm exp}_{\mathrm stat}=1496\pm 3\ ({\mathrm stat.})
\pm 3\ ({\mathrm syst.})\ {\mathrm s}^{-1}=
1496\pm 4\ {\mathrm s}^{-1}\ \ \ .
\label{eq:lexp}
\end{equation}
This result, which confirms earlier experiments of much less
statistics\cite{3He65}, is in beautiful agreement with the latest 
and most precise theoretical analysis\cite{Con92} whose prediction is
$\lambda^{\mathrm theor}_{\mathrm stat}=1497\pm 21\ {\mathrm s}^{-1}$.

Given such precision measurements and prospects for new ones using polarised
\nuc{3}{He} targets, it thus seems timely to consider the possible
physics information which may be gathered from such experiments.
In the next section, this issue is first addressed within the Standard
Model (SM) for the electroweak interactions. The following section
considers the sensitivity of the aforementioned observables to possible
new interactions which may be lurking behind the horizon of the
Standard Model. In both cases, the complementarity of different types
of measurements is most apparent, thus calling for new
experiments aiming for the best precisions attainable, as well as for
a renewed theoretical effort to improve on the precision of different
inputs.

Even though the present analysis is developed in the context of muon 
capture on \nuc{3}{He}, a similar discussion applies equally well
in the case of muon capture on hydrogen. Indeed, both cases being that
of muon capture on a spin $\half$ isospin doublet, the relevant expressions
remain identical; only numerical values of form factors differ from
one case to the other. In fact, the most general situation possible
is being analysed\cite{Lucio}---including polarised muons and 
\nuc{3}{He} nuclei, both within the Standard Model and beyond it---in order
to ascertain the interest of measurements other than the statistical
capture rate, be it the triton asymmetry or new experiments altogether such
as the triton polarisation or tensor analysing power.
Some of the present conclusions are based on preliminary results 
of that work in progress.

\section{Muon Capture in the Standard Model}
\label{Sect2}

Within the SM, the charged current interaction possesses the $(V-A)$
left-handed chirality structure in both the muonic and hadronic vertices.
The effective interaction amplitude is of the form
$\frac{g^2}{8M^2}\,V_{ud}\,{J^\mu_{\mathrm lept}}^\dagger\ 
J_{{\mathrm hadr}\,\mu}$,
with $g$ being the usual SU(2)$_{\mathrm L}$ gauge coupling constant, $M$
the ${\mathrm W}^\pm$ gauge boson mass, 
and $V_{ud}$ the Cabibbo-Kobayashi-Maskawa
flavour mixing matrix element. The leptonic current is 
$J^\mu_{\mathrm lept}=\bar{\mu}\gamma^\mu(1-\gamma_5)\nu_\mu$, the
neutrino being massless without flavour mixing.
The hadronic current is of the form
$J^\mu_{\mathrm hadr}=V^\mu_{\mathrm hadr}-A^\mu_{\mathrm hadr}$, with the
vector and axial contributions given by the general Lorentz covariant 
parametrisation in momentum space,
\begin{equation}
\begin{array}{r c l}
V^\mu_{\mathrm hadr}&=&\overline{\psi}_2\,
\Big[\,F_V\gamma^\mu+iF_M\sigma^{\mu\nu}\frac{q_\nu}{2M}+F_S\frac{q^\mu}{2M}\,
\Big]\psi_1\ \ \ ,\\
A^\mu_{\mathrm hadr}&=&\overline{\psi}_2\,
\Big[\,F_A\gamma^\mu\gamma_5+F_P\gamma_5\frac{q^\mu}{2M}+
iF_T\sigma^{\mu\nu}\frac{q_\nu}{2M}\, \Big]\psi_1\ \ \ .\\
\end{array}
\label{eq:VA}
\end{equation}
Here, $\psi_1$ and $\psi_2$ are Dirac spinors for on mass-shell 
initial and final spin~$\half$ nuclei of energy-momenta $p_1$ and $p_2$,
respectively, $M$ is a mass scale chosen
to be their average mass value, and $q^\mu=p_2^\mu-p_1^\mu$ 
is the energy-momentum transfer of the process. 
The quantities $F_V$, $F_M$, $F_S$, $F_A$, $F_P$ and
$F_T$ are phenomenological $q^2$-dependent nuclear form factors, 
real under complex conjugation for time reversal invariant interactions,
whose values may be determined using experimental data and the 
hadronic symmetries of CVC and PCAC\footnote{In the case of muon capture
on hydrogen, the corresponding form factors are usually denoted by
$g_V$, $g_M$, $g_S$, $g_A$, $g_P$ and $g_T$, respectively. The values of
$|g_S/g_V|$ and $|g_T/g_A|$ should not exceed, say, 0.02, on account
of isospin symmetry breaking\cite{Shio96}.}. 
In the limit of exact isospin symmetry
and charge conjugation invariance, the induced second-class form factors 
$F_S$ and $F_T$ vanish identically; by analogy with the si\-tu\-ation for
the nucleon\cite{Shio96}, their normalised values $|F_S/F_V|$
and $|F_T/F_A|$ should thus not exceed, say, 0.02.
The vector form factors $F_V$ and $F_M$ are rather well established on the
basis of CVC, taking the following values\cite{Con92} at the relevant invariant 
momentum transfer $q^2_0=-0.954\,m^2_\mu$:
$F_V(q^2_0)=0.834\pm 0.011$ and $F_M(q^2_0)=-13.969\pm 0.052$.
The axial form factor $F_A(q^2=0)$ is well determined from the $\beta$-decay
rate of \nuc{3}{H}, but its extrapolation to $q^2_0$ requires some
model dependent assumptions. In Ref.\cite{Con92}, two uncertainties
are considered for that extrapolation, the larger of which accounts 
generously for meson exchange contributions possibly overlooked in the 
smaller one, even though the latter should already be rather reliable since it
uses the $q^2$-dependence of $F_A$ inferred from the impulse 
approximation\footnote{A precise calculation of the capture rate
is also available in the impulse approximation including meson exchange
currents\cite{Con96}, which,
on basis of the experimental result (\ref{eq:lexp}), enables
a test\cite{Con96,Acker97} of the QCD-corrected PCAC 
prediction\cite{Bern94} for $g_P$, the precision of the latter being
better than 3\%.}.
Correspondingly, the value for $F_A$ is\cite{Con92}
$F_A(q^2_0)=-1.052\pm 0.005\ ({\mathrm or} \pm 0.010)$.
Finally, the value for the induced pseudoscalar form factor $F_P$ may
be related to that of $F_A$ on the basis of PCAC---a consequence of the
spontaneous breaking of the approximate chiral symmetries of QCD---, through
$F^{\mathrm PCAC}_P(q^2)=4MF_A(q^2)/(m^2_\pi-q^2)$.
Here, a correction---not exceeding a few percent\cite{Klie84}---due 
a possibly different $q^2$-dependence of the
$\pi$-\nuc{3}{He}-\nuc{3}{H} coupling constant and of $F_A$ is ignored.

Given the above parametrisation of the muon capture amplitude, it is
possible to compute different 
observables\cite{Kim65,Hwang78,Con92}\footnote{Refs.\cite{Kim65,Hwang78,Con92} 
ignore the induced second-class contributions, 
while Ref.\cite{Sant76} includes them
but only numerically. Ref.\cite{Lucio} will provide explicit analytic 
expressions in the most general case possible.},
in particular the statistical capture rate $\lambda_{\mathrm stat}$ and the
triton asymmetry $A_{\mathrm v}$. In the limit that both 
$F_S$ and $F_T$ vanish, and with the values quoted above for the other 
form factors, Ref.\cite{Con92} obtains 
$\lambda^{\mathrm theor}_{\mathrm stat}=
1497\pm 12\ ({\mathrm or} \pm 21)\ {\mathrm s}^{-1}$ and
$A^{\mathrm theor}_{\mathrm v}=0.524\pm 0.006\ ({\mathrm or} \pm 0.006)$,
where the first (resp. second) error indicated in each case 
corresponds to the smallest (resp. largest) uncertainty of $\pm 0.005$ 
(resp. $\pm 0.010$) on $F_A$ (this notation is used throughout). 
Note how $A_{\mathrm v}$ is much less sensitive
to $F_A$ than is $\lambda_{\mathrm stat}$. Indeed, the normalised
variations $F_A/{\cal O}\,d{\cal O}/dF_A$, with 
$({\cal O}=\lambda_{\mathrm stat},A_{\mathrm v}$), are $1.521$ and $-0.134$,
respectively. The situation is quite the opposite with respect
to $F_P$, since the normalised variations are then\cite{Con92}
$-0.116$ and $-0.377$, respectively. Consequently, a precision measurement
of $A_{\mathrm v}$ provides a determination of $F_P$ not much sensitive to
the uncertainty on $F_A$, in contradistinction to $\lambda_{\mathrm stat}$.
Moreover, given the values for $F_V$ and $F_M$ and assuming no second-class
contributions $F_S=0=F_T$, a combined analysis using these two observables
would lead to a purely experimental determination both of $F_A$ and of $F_P$,
independent of the \nuc{3}{H} $\beta$-decay rate and of assumptions on
the $q^2$-dependence of $F_A$.

To numerically assess the situation when $F_S=0=F_T$, 
let us consider the experimental
result $\lambda^{\mathrm exp}_{\mathrm stat}$ in (\ref{eq:lexp}), as well as
a hypothetical value for $A_{\mathrm v}$ equal to its
prediction $A^{\mathrm theor}_{\mathrm v}=0.524$ with a relative precision of
5\% and 1\%. From $\lambda^{\mathrm exp}_{\mathrm stat}$, one 
obtains\cite{Acker97},
\begin{equation}
F^{\mathrm norm}_P\equiv \frac{m_\mu}{2M}F_P(q^2_0)=-20.80\pm 1.57\ 
({\mathrm or} \pm 2.74)\ \ \ ,
\label{eq:Fnorm}
\end{equation}
to be compared to the
PCAC prediction of $-20.72\pm 0.10$ (or $\pm 0.20$). This $\pm 7.6\%$
(or $\pm 13.2\%$) result is thus a beautiful confirmation 
of PCAC\footnote{In turn, this test of PCAC allows both for a somewhat 
improved\cite{Con96,Acker97} precision of 
$g_P=(1.05\pm 0.19)g^{\mathrm PCAC}_P$
on the present combined 22\% uncertainty on $g_P$\cite{Bard81} from all muon
capture measurements on hydrogen, and for
a precise determination of the pion-nucleus coupling constant\cite{Muk96}.},
in contradistinction with the conclusion reached recently
from a radiative muon capture experiment on hydrogen\cite{Jonk96} which
finds a value for $g_P$ 1.5 times larger than the PCAC prediction\cite{Bern94}.
On the other hand, a 5\% (resp. 1\%) precision on $A_{\mathrm v}$ would
lead to uncertainties on $F^{\mathrm norm}_P$ of $\pm 2.81$ (resp. $\pm 0.80$),
irrespective of whether the error on $F_A$ is $\pm 0.005$ or $\pm 0.010$. 
Therefore, in order to improve the precision on $F^{\mathrm norm}_P$ 
reached using (\ref{eq:lexp}), the precision aim of a triton asymmetry 
measurement should at least be on the order of 1\% to 2\%, a real 
challenge indeed. 
It may also be shown that for a combined analysis using
both $\lambda_{\mathrm stat}$ and $A_{\mathrm v}$,
and when accounting for the errors both
on $\lambda^{\mathrm exp}_{\mathrm stat}$ and $A^{\mathrm exp}_{\mathrm v}$,
and on $F_V$ and $F_M$, a 1\% measurement
of $A_{\mathrm v}$ would lead to a determination\footnote{The
precision on $F_P$ is then set by that on $A_{\mathrm v}$, namely
$\pm 0.80$.} of $F_A$ with an
uncertainty close to the value of $\pm 0.005$.
As a matter of fact, when the errors on $F_V$, $F_M$ and $F_A$ are not 
included but only those on the experimental values 
$\lambda^{\mathrm exp}_{\mathrm stat}$ and
$A^{\mathrm exp}_{\mathrm v}$, the present uncertainty
on $F^{\mathrm norm}$ as determined from (\ref{eq:lexp}) is reduced
to $\pm 0.48$, while using the value for
$A_{\mathrm v}$ precise to 5\% (resp. 1\%)
brings this uncertainty down to $\pm 2.75$ (resp. $\pm 0.55$). This establishes
that on the theoretical front as well, an improvement on the determination
of the form factors $F_V$, $F_M$ and $F_A$ is clearly also called for,
especially in the case of $F_A$ whose error presently dominates the uncertainty
on $F^{\mathrm norm}_P$ given above.

A similar analysis is possible\cite{Hol84} for the second-class 
form factors $F_S$ and
$F_T$, when assuming that either one of these vanishes, while the remaining
form factors take the values listed above, with $F_P$ given by the PCAC
relation quoted previously. An analysis using both 
$\lambda_{\mathrm stat}$ and $A_{\mathrm v}$ could also determine $F_S$
and $F_T$ from a combined fit. Here again, the sensitivity of 
$A_{\mathrm v}$ to $F_S$ and $F_T$
is much larger than that of $\lambda_{\mathrm stat}$, since
with $({\cal O}=\lambda_{\mathrm stat},A_{\mathrm v})$, one
has $1/{\cal O}\,d{\cal O}/dF_S=(0.007,0.017)$ and
$1/{\cal O}\,d{\cal O}/dF_T=(-0.006,-0.019)$.
The experimental result (\ref{eq:lexp}) then implies,
\begin{equation}
F_S=-0.062\pm 1.18\ ({\mathrm or} \pm 2.02)\ \ ,\ \ 
F_T=0.075\pm 1.43\ ({\mathrm or} \pm 2.45)\ \ ,
\label{eq:FSFT}
\end{equation}
while these uncertainties reduce to $\pm 0.38$ and $\pm 0.46$, 
respectively, when only
the experimental error on $\lambda^{\mathrm exp}_{\mathrm stat}$ is
included. These results thus provide some improvement on the uncertainties
obtained in Ref.\cite{Hol84} in the impulse appro\-xi\-mation for the nucleon
form factors $g_S$ and $g_T$, but with central values much closer to
expectations. Nevertheless, they also call for a theoretical
improvement on the uncertainties of $F_V$, $F_M$ and especially of $F_A$,
for the same reasons as above.

Similarly, a measurement of $A_{\mathrm v}$ precise
to 5\% (resp. 1\%) at $A^{\mathrm theor}_{\mathrm v}=0.524$ would imply
uncertainties on $F_S$ and $F_T$ of $\pm 3.0$ (resp. $\pm 0.9$)
and $\pm 2.8$ (resp. $\pm 0.8$), respectively, whether the error on 
$F_A$ is $\pm 0.005$ or $\pm 0.010$. 
When only the experimental uncertainty on $A_{\mathrm v}$ is included, these
numbers reduce to $\pm 2.9$ (resp. $\pm 0.58$) and $\pm 2.7$
(resp. $\pm 0.54$), respectively. The situation is thus comparable to
that for $F_P$. When accounting for the present errors on $F_V$, $F_M$ 
and $F_A$, a 1\% measurement of $A_{\mathrm v}$ would improve
by about a factor two the present uncertainties both on $F_S$ and $F_T$
established in (\ref{eq:FSFT}) on basis of the experimental result
(\ref{eq:lexp}). Such results would also be comparable or would
improve some of the limits presently available 
on these contributions from $\beta$-decay processes\cite{Gren85}.

\section{Muon Capture Beyond the Standard Model}
\label{Sect3}

Assuming the given values for the hadronic form factors\footnote{In this
first order analysis, the possible second-class contributions of $F_S$ 
and $F_T$ are both set equal to zero, an approximation which, in view of the
results of Sect.\ref{Sect2}, is already quite satisfactory. Realistic
values for $F_S$ and $F_T$ would mostly affect only the uncertainties
on the constraints derived in the present section, and then by a small
amount at best.}, precision measurements of $\lambda_{\mathrm stat}$ and 
$A_{\mathrm v}$ may also be used to set limits for physics beyond the SM.
At the $(u,d)$ quark level, such new interactions may effectively 
be represented by the amplitude,
\begin{equation}
\begin{array}{r l}
&4\frac{g^2}{8M^2}V_{ud}\sum_{\eta_1,\eta_2=\pm,\pm}\,\Bigg[\ 
\left(h^V_{\eta_1\eta_2}\right)^*\bar{\nu}_\mu\gamma^\mu P_{\eta_1}\mu\,
\bar{d} \gamma_\mu P_{\eta_2}u\,+\,\\
&\,+\left(h^S_{\eta_1\eta_2}\right)^*\bar{\nu}_\mu P_{\eta_1}\mu\,
\bar{d} P_{\eta_2}u\,+\,
\left(h^T_{\eta_1\eta_2}\right)^*\bar{\nu}_\mu\sigma^{\mu\nu} P_{\eta_1}\mu\,
\bar{d} \sigma_{\mu\nu} P_{\eta_2}u\ \Bigg]\ \ \ ,
\end{array}
\label{eq:effec}
\end{equation}
where $P_\pm=(1\pm\gamma_5)/2$ are the chirality projectors, and $g$,
$M$ and $V_{ud}$ are arbitrary parameters which in the limit of the SM
reduce to the usual ones introduced in Sect.\ref{Sect2}. Finally, the
coefficients $h^{V,S,T}_{\pm\pm}$ are arbitrary complex coefficients
associated to vector, scalar and tensor interactions in the charge
exchange form, with $\eta_1$ (resp. $\eta_2$) being the muon 
(resp. $u$ quark) chirality which equals the neutrino (resp. $d$ quark) 
chirality for vector interactions and is opposite to it for scalar 
and tensor interactions.
Finally, without loss of generality, one may set $h^T_{--}=0=h^T_{++}$.
The SM amplitude is recovered in the limit that all coefficients
$h^{S,V,T}_{\pm,\pm}$ vanish except for $h^V_{--}=+1$.

Like-wise, effective interactions for $\beta$- and $\mu$-decay 
may be parametrised in terms of analogous coefficients 
$f^{V,S,T}_{\pm\pm}$ and $g^{V,S,T}_{\pm\pm}$, respectively, with
the electron playing the r\^ole of the muon in the two cases,
and the muon playing that of the $d$ quark in the latter case\footnote{In the
case of muon decay, this parametrisation is now standard\cite{PDG}.}. 
Consequently, the relation between
the coupling constant $g^2/8M^2$ and Fermi's constant $G_F$, as well as the
mixing parameter $V_{ud}$, determined from the muon decay rate and
$0^+-0^+$ superallowed $\beta$-decay rates, respectively, 
are modified accordingly, leading to a computable correction factor for the 
muon capture rate. Note that the amplitude in (\ref{eq:effec}) also
determines the $\pi\rightarrow\mu\nu_\mu$ decay rate, as well as the
$\pi\rightarrow e\nu_e$ decay rate in terms of the coefficients 
$f^{S,V,T}_{\pm\pm}$. In these two cases however, only the axial and
pseudoscalar combinations of coefficients contribute. Hence, the
experimental branching ratio 
$R^{\mathrm exp}_\pi=\Gamma(\pi^+\rightarrow e^+\nu_e(\gamma))/
\Gamma(\pi^+\rightarrow \mu^+\nu_\mu(\gamma))$ corrected for
radiative corrections such that\cite{Herc95}
$R^{\mathrm exp}_\pi/R^{\mathrm SM}_\pi=0.9960\pm 0.0033$ implies
the constraint,
\begin{equation}
\begin{array}{r l}
&\frac{|(h^V_{--}-h^V_{-+})+C_\mu(h^S_{++}-h^S_{+-})|^2+
|(h^V_{+-}-h^V_{++})+C_\mu(h^S_{-+}-h^S_{--})|^2}
{|(f^V_{--}-f^V_{-+})+C_e(f^S_{++}-f^S_{+-})|^2+
|(f^V_{+-}-f^V_{++})+C_e(f^S_{-+}-f^S_{--})|^2}=\\
&=1.0040\pm 0.0033\ \ \ ,
\end{array}
\label{eq:pidecay}
\end{equation}
with $C_\mu=m^2_\pi/(m_\mu(m_u+m_d))\simeq 12.29$ and 
$C_e=m^2_\pi/(m_e(m_u+m_d))\simeq 2541$.
Note that it is only when the interaction is a purely left-handed vector one
that this constraint provides a $e-\mu$ universality test of the SM, namely
$|h^V_{--}/f^V_{--}|^2=1.0040\pm 0.0033$.

In the case of muon capture on \nuc{3}{He}, let us also first consider
this situation, namely when all coefficients $h^{S,V,T}_{\pm\pm}$
and $f^{S,V,T}_{\pm\pm}$ vanish except for $h^V_{--}$ and $f^V_{--}$
which are left unspecified. The experimental result (\ref{eq:lexp})
for the capture rate then implies the constraint,
\begin{equation}
\Big|\frac{h^V_{--}}{f^V_{--}}\Big|^2=0.9996\pm 0.0083\
({\mathrm or} \pm 0.0142)\ \ \ ,
\label{eq:univ}
\end{equation}
the uncertainty on this constraint reducing to $\pm 0.0027$ when only
the experimental error on $\lambda_{\mathrm stat}$ is included.
Therefore, under these much restricted assumptions, and because of
the uncertainties on the form factors---mostly that on $F_A$---, this
$e-\mu$ universality test does not improve on the present limit
from pion decay. Nevertheless, once the parameter space is enlarged,
the two types of observables lead to complementary constraints on the
effective coupling coefficients $h^{S,V,T}_{\pm\pm}$ and $f^{S,V,T}_{\pm\pm}$.

General left-right symmetric models (LRSM)\cite{Moh86} 
whose gauge symmetry group 
SU(2)$_{\mathrm L}$$\times$SU(2)$_{\mathrm R}$$\times$U(1)$_{\mathrm B-L}$ 
provides an attractive gauge extension of the SM, are an interesting
context for the present discussion. Let us introduce the notations,
\begin{equation}
\begin{array}{r l}
&r=\frac{g_R}{g_L}\ \ ,\ \ \delta=\frac{M^2_1}{M^2_2}\ \ ,\ \ 
t=\tan\zeta\ \ ,\\
&v_{ud}=\frac{V^R_{ud}}{V^L_{ud}}\ \ ,\ \ v_u=|v_{ud}|^2\ \ ,\ \ 
v_\ell=\frac{\sum_i'|U^R_{\ell i}|^2}{\sum_i'|U^L_{\ell i}|^2}\ \ ,\ \ 
\ell=e,\mu\ \ .
\end{array}
\end{equation}
Here, $g_R$ and $g_L$ are the gauge coupling constants of the chiral
SU(2)$_{\mathrm R,L}$ gauge groups, $M_1$ and $M_2$ are the masses of the 
physical charged gauge bosons ${\mathrm W}^\pm_1$ and 
${\mathrm W}^\pm_2$---the former being the lightest---, $\zeta$ is their 
mixing angle in terms of the ${\mathrm W}^\pm_{L,R}$ gauge bosons, 
and $V^{R,L}_{ud}$, $U^{R,L}_{\ell i}$ are Cabibbo-Kobayashi-Maskawa flavour 
mixing matrix elements in each chiral sector, both in the $(u,d)$ quark 
sector and in the lepton sector, the latter coupling the charged lepton of 
flavour $\ell$ to the mass eigenstate neutrino $\nu_i$. The prime ``$\,'\,$" 
on the summation symbols in the lepton sector indicates that only those 
neutrinos whose mass is sufficiently small in order to be produced in a 
given process are included, while numerically in this note
their mass is then also neglected in the 
kinematics of that process. Finally, in the mixing of the 
${\mathrm W}^\pm_{R,L}$ gauge bosons leading 
to the physical charged bosons ${\mathrm W}^\pm_{1,2}$, 
in general there may also appear a CP violating phase $\omega$.  
By definition, so-called manifest left-right symmetric models (MLRSM)
are such that $r=1$, $v_{ud}=1$, $v_\mu=1=v_e$ and $\omega=0$.

Within LRSM, and when the much suppressed charged Higgs contributions
are ignored, only effective vector interactions are induced, leading
in general to non vanishing coefficients $h^V_{\pm\pm}$ and $f^V_{\pm\pm}$.
First, let us assume that the neutrinos produced in $\mu^--$\nuc{3}{He} capture 
and in $0^+-0^+$ superallowed $\beta$-decays are such that $v_\mu=v_e$, and
let us also take it for granted that $(\tan\zeta)$ and $\delta$ 
are sufficiently small to warrant an expansion in these two parameters.
Under these circumstances, the muon capture rate $\lambda_{\mathrm stat}$ is 
essentially independent\footnote{This is also the case for
the hyperfine muon capture rates.} of the mass $M_2$ of the heavier
charged gauge boson, but is rather sensitive\cite{BLYP58} to the 
mixing angle $\zeta$, while the situation is reversed for the triton asymmetry 
$A_{\mathrm v}$. Given these assumptions, the experimental value 
(\ref{eq:lexp}) for the capture rate then leads to the result,
\begin{equation}
\frac{g_R}{g_L}\,{\mathrm Re}\left(e^{i\omega}\,v_{ud}\right)\,\tan\zeta=
-0.00016\pm 0.00295\ ({\mathrm or} \pm 0.00506)\ \ \ ,
\end{equation}
with the uncertainty reducing to $\pm 0.00095$ when only the experimental
error on $\lambda^{\mathrm exp}_{\mathrm stat}$ is included. This
new constraint on the mixing angle $\zeta$, which applies in general
LRSM and thus includes a dependence on $r$ and $e^{i\omega}v_{ud}$, 
is among the very best available in the context of MLRSM
for which all other such constraints have been obtained from muon decay,
$\beta$-decay and other semi-leptonic processes\cite{Bern96,Deut95,Mix}.

Given that the mixing angle $\zeta$ must be small\footnote{Unless 
$v_{ud}$ vanishes altogether, which would make the collider
limits\cite{WR} on $M_2$ void of content.}, let us now relax
the condition $v_\mu=v_e$ but take the limit $\zeta=0$.
Corresponding to the universality limit (\ref{eq:univ}),
the experimental result (\ref{eq:lexp}) then implies the
constraint\footnote{Here, $v_\mu$ includes the sum over all neutrinos
produced in muon capture, while $v_e$ includes all neutrinos produced
in $0^+-0^+$ superallowed $\beta$-decays.},
\begin{equation}
(v_\mu-v_e)\,v_u\,r^4\,\delta^2=0.0004\pm 0.0083\ 
({\mathrm or} \pm 0.0142)\ \ \ ,
\end{equation}
the uncertainty reducing to $\pm 0.0027$ when only the experimental
error on $\lambda^{\mathrm exp}_{\mathrm stat}$ is included.
{}From pion decay, the analogous constraint is 
$(v^{(\pi)}_\mu-v^{(\pi)}_e)v_u r^4\delta^2=0.0040\pm 0.0033$,
where $v^{(\pi)}_\mu$ and $v^{(\pi)}_e$ include summations over all
neutrinos produced in the muon and electron branches
of pion decay, respectively.

Finally, $A_{\mathrm v}$ being essentially insensitive to $\zeta$,
in the approximation that $\zeta=0$ one obtains\cite{Gov95},
\begin{equation}
\left(A^{\mathrm LRSM}_{\mathrm v}\right)_{|\zeta=0}=
\frac{1-v_\mu\,v_u\,r^4\,\delta^2}{1+v_\mu\,v_u\,r^4\,\delta^2}\
A^{\mathrm SM}_{\mathrm v}\ \ \ .
\end{equation}
Therefore, a 1\% measurement of $A_{\mathrm v}$ provides an upper bound 
of 0.005 on $|v_\mu v_u r^4\delta^2|$, which in the case of MLRSM
corresponds to a lower bound on $M_2$ of 300 GeV/c$^2$ only.

It is also interesting to consider limits
that the experimental result (\ref{eq:lexp}) may set on scalar
and tensor interactions, characterised by the coefficients $h^S_{\pm\pm}$
and $h^T_{\pm\pm}$, which typical arise in theories with leptoquarks. 
Mostly for illustrative purposes, only interactions
coupling to left-handed neutrinos are included here, motivated by the
idea that right-handed ones may be too heavy to be produced in muon capture
and $\beta$-decay.
This corresponds to $h^V_{+\pm}=h^S_{-\pm}=h^T_{-\pm}=0$, and
similarly for the $f^{S,V,T}_{\pm\pm}$ coefficients.
Moreover, for simplicity, one also assumes that
$|h^V_{-+}+h^V_{--}|^2\simeq |f^V_{-+}+f^V_{--}|^2$.
Associated to such interactions, one must also consider the corresponding
hadronic matrix elements of the $(u,d)$ scalar, pseudoscalar and tensor
currents $\bar{d}u$, $\bar{d}\gamma_5 u$ and $\bar{d}\sigma^{\mu\nu}u$,
respectively. In the same way as in (\ref{eq:VA}) for the vector and
axial currents, these matrix elements define the scalar, pseudoscalar
and tensor nuclear form factors $G_S$, $G_P$ and $G_T$, respectively,
where in contradistinction to (\ref{eq:VA}), induced recoil corrections
are not included for the obvious reason that they would be even further
suppressed in comparison to the genuine scalar, pseudoscalar and tensor
interactions.

First, consider the case when the only such additional interactions are 
restricted to be the scalar ones, namely when $h^T_{+-}=0$, and assume then 
also that\footnote{This condition is certainly satisfied to a very good 
approximation.}
$|G^{(0)}_S|^2|f^S_{++}+f^S_{+-}|^2<<
|F^{(0)}_V|^2|f^V_{-+}+f^V_{--}|^2$, where $G^{(0)}_S$ and $F^{(0)}_V$ are the
hadronic matrix elements in the case of $0^+-0^+$ superallowed $\beta$-decays
for the scalar and vector $(u,d)$ quark currents, respectively.
Then, when either one of the pseudoscalar or scalar combinations
$(h^S_{++}-h^S_{+-})$ and $(h^S_{++}+h^S_{+-})$
vanishes, the experimental value (\ref{eq:lexp}) leads to the
contraints, respectively,
\begin{equation}
\begin{array}{r l}
\frac{h^S_{++}+h^S_{+-}}{h^V_{--}+h^V_{-+}}\,G_S&=
-0.0012\pm 0.0221\ ({\mathrm or} \pm 0.0380)\ \ ,\ \ \\
\frac{h^S_{++}-h^S_{+-}}{h^V_{--}+h^V_{-+}}\,G_P&=
-0.08\pm 1.49\ ({\mathrm or} \pm 2.56)\ \ .\ \ 
\end{array}
\label{eq:SPS}
\end{equation}
When only the experimental error on $\lambda^{\mathrm exp}_{\mathrm stat}$ 
is included, these uncertainties reduce to $\pm 0.0071$ and $\pm 0.480$, 
respectively. In fact\cite{Lucio}, the result involving $G_S$ 
is the one obtained in (\ref{eq:FSFT}) for $m_\mu F_S/(2M)$, while 
the precision on the result involving $G_P$ is that obtained for
$F^{\mathrm norm}_P$ in (\ref{eq:Fnorm}). Clearly, these limits are
complementary to those which, under the same approximations, may be obtained
from the pion decay branching ratio (\ref{eq:pidecay}) for the pseudoscalar 
couplings $(f^S_{++}-f^S_{+-})$ and $(h^S_{++}-h^S_{+-})$, 
given the rather large ratio $C_e/C_\mu\simeq 207$. In this respect, the
stringent limit (\ref{eq:SPS}) for the scalar coupling $(h^S_{++}+h^S_{+-})$ is
quite interesting.

Still under the same chirality assumptions as stated above in relation to
left-handed neutrinos, consider now the situation when only new tensor 
interactions are included, namely when $h^S_{++}=0=h^S_{+-}$. In that case,
(\ref{eq:lexp}) provides the constraint,
\begin{equation}
h^T_{+-}\,G_T=-0.00008\pm 0.00143\ ({\mathrm or} \pm 0.00245)\ \ \ ,
\end{equation}
the uncertainty reducing to $\pm 0.000460$ when only the experimental
error on $\lambda^{\mathrm exp}_{\mathrm stat}$ is included.
This result thus determines from the muon capture rate on \nuc{3}{He}
a most stringent upper bound on tensor interactions coupling to 
left-handed neutrinos only.

Limits that may be set similarly on scalar and tensor interactions
from a 1\% measurement of the triton asymmetry $A_{\mathrm v}$ will
be considered elsewhere\cite{Lucio}. In view of the results obtained
in that case in Sect.\ref{Sect2} for the second-class induced form 
factors $F_S$ and $F_T$, one may expect quite stringent constraints 
to apply as well.

\section{Conclusions}

As this contribution illustrates, precision measurements in muon
capture on \nuc{3}{He} to the triton channel provide for stringent
tests of the charged electroweak interaction, both within the Standard Model
and for new physics beyond it. The recent precision result\cite{Acker97}
for the statistical capture rate enables a series of such tests, some new,
the others improving on previous limits, which are complementary to
tests that precision measurements of other observables using polarised
\nuc{3}{He} targets would make possible. A 1\% measurement of the triton
asymmetry\cite{Cum95} is certainly such an example offering much interest,
but the potential of other possibilities should also be 
assessed\cite{Lucio}. Making the most
out of muon capture on polarised \nuc{3}{He} calls both for much dedicated
experimental efforts and ingenuity, as well as for a renewed theoretical
evaluation of the inputs involved in the description of that process.

\section*{Acknowledgements}

It is a pleasure to thank Prof. J.~Deutsch for many valuable discussions.
My thanks also go to Prof. M.~Tanaka
for his kind invitation to a most interesting Workshop.

\end{document}